# Controlling Electromagnetic Fields

**J. B. Pendry**[1], **D. Schurig**[2] **and D. R. Smith**[2]
[1]Department of Physics, The Blackett Laboratory, Imperial College London, London SW7 2AZ, UK,

[2]Department of Electrical and Computer Engineering, Duke University, Box 90291, Durham, NC 27708, USA.

Using the freedom of design which metamaterials provide, we show how electromagnetic fields can be redirected at will and propose a design strategy. The conserved fields: electric displacement field, **D**, magnetic induction field, **B**, and Poynting vector, **S**, are all displaced in a consistent manner. A simple illustration is given of the cloaking of a proscribed volume of space to exclude completely all electromagnetic fields. Our work has relevance to exotic lens design and to the cloaking of objects from electromagnetic fields.

To exploit electromagnetism we use materials to control and direct the fields: a glass lens in a camera to produce an image, a metal cage to screen sensitive equipment, 'black bodies' of various forms to prevent unwanted reflections. With homogeneous materials, optical design is largely a matter of choosing the interface between two materials. For example, the lens of a camera is optimized by altering its shape so as to minimize geometrical aberrations. Electromagnetically inhomogeneous materials offer a different approach to control light; the introduction of specific gradients in the refractive index of a material can be used to form lenses and other optical elements, although the types and ranges of such gradients tend to be limited.

A new class of electromagnetic materials (1,2) is currently under study: metamaterials, which owe their properties to sub-wavelength details of structure rather than to their chemical composition, can be designed to have properties difficult or impossible to find in nature. In this report we show how the design flexibility of metamaterials can be exploited to achieve new and remarkable electromagnetic devices. The message of this paper is that metamaterials enable a new paradigm for the design of electromagnetic structures at all frequencies from optical down to DC.

Progress in the design of metamaterials has been impressive. A negative index of refraction (3) is an example of a material property that does not exist in nature, but has been enabled using metamaterial concepts. As a result, negative refraction has been much studied in recent years (4) and realizations have been reported at both GHz and optical frequencies (5-8). Novel magnetic properties have also been reported over a wide spectrum of frequencies. Further information on the design and construction of metamaterials may be found in (9-13). In fact, it is now conceivable that a material can be constructed whose permittivity and permeability values may be designed to vary independently and arbitrarily throughout a material, taking positive or negative values as desired

If we take this unprecedented control over the material properties and form inhomogeneous composites, we enable a new and powerful form of electromagnetic design. As an example of this design methodology, we show how the conserved quantities of electromagnetism: the electric displacement field, **D**, the magnetic field intensity, **B**, and the Poynting vector, **S**, can all be directed at will, given access to the appropriate metamaterials. In particular these fields can be focused as required or made to avoid objects and flow around them like a fluid, returning undisturbed to their original trajectories. These conclusions follow from exact manipulations of Maxwell's equations and are not confined to a ray approximation. They encompass in principle all forms of electromagnetic phenomena on all length scales.

We start with an arbitrary configuration of sources embedded in an arbitrary dielectric and magnetic medium. This initial configuration would be chosen to have the same topology as the final result we seek. For example, we might start with a uniform electric field and require that the field lines be moved to avoid a given region. Next imagine that the system is embedded in some elastic medium that can be pulled and stretched as we desire (Fig. 1). To keep track of distortions we record the initial configuration of the fields on a Cartesian mesh which is subsequently distorted by the same pulling and stretching process. The distortions can now be recorded as a coordinate transformation between the original Cartesian mesh and the distorted mesh:

$$u(x,y,z), v(x,y,z), w(x,y,z) \qquad (1)$$

where (u, v, w) is the location of the new point with respect to the x, y, z axes. What happens to Maxwell's equations when we substitute the new coordinate system? The equations have exactly the same form in any coordinate system, but the refractive index— or more exactly the permittivity, ε, and permeability, μ —are scaled by a common factor. In the new coordinate system we must use renormalized values of the permittivity and permeability:

$$\varepsilon'_u = \varepsilon_u \frac{Q_u Q_v Q_w}{Q_u^2},$$
$$\mu'_u = \mu_u \frac{Q_u Q_v Q_w}{Q_u^2}, \quad \text{etc.} \qquad (2)$$

$$E'_u = Q_u E_u, \quad H'_u = Q_u H_u, \quad \text{etc.} \qquad (3)$$

where,

$$Q_u^2 = \left(\frac{\partial x}{\partial u}\right)^2 + \left(\frac{\partial y}{\partial u}\right)^2 + \left(\frac{\partial z}{\partial u}\right)^2$$

$$Q_v^2 = \left(\frac{\partial x}{\partial v}\right)^2 + \left(\frac{\partial y}{\partial v}\right)^2 + \left(\frac{\partial z}{\partial v}\right)^2 \qquad (4)$$

$$Q_w^2 = \left(\frac{\partial x}{\partial w}\right)^2 + \left(\frac{\partial y}{\partial w}\right)^2 + \left(\frac{\partial z}{\partial w}\right)^2$$

As usual,

$$\mathbf{B}' = \mu_0 \boldsymbol{\mu}' \mathbf{H}', \qquad \mathbf{D}' = \varepsilon_0 \boldsymbol{\varepsilon}' \mathbf{E}' \qquad (5)$$

We have assumed orthogonal coordinate systems for which the formulae are particularly simple. The general case is given in (13) and in the accompanying online material. The equivalence of coordinate transformations and changes to ε and μ has also been referred to in (14).

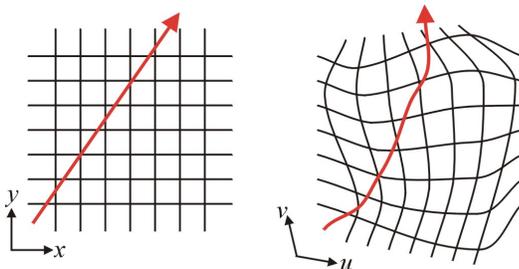

Fig. 1 Left: a field line in free space with the background Cartesian coordinate grid shown. Right: the distorted field line with the background coordinates distorted in the same fashion. The field in question may be the electric displacement or magnetic induction fields **D**, **B**, or the Poynting vector, **S**, which is equivalent to a ray of light.

Now let us put these transformations to use. Suppose we wish to conceal an arbitrary object contained in a given volume of space; furthermore, we require that external observers be unaware that something has been hidden from them. Our plan is to achieve concealment by cloaking the object with a metamaterial whose function is to deflect the rays that would have struck the object, guide them around the object, and return them to their original trajectory.

Our assumptions imply that no radiation can get into the concealed volume, nor can any radiation get out. Any radiation attempting to penetrate the secure volume is smoothly guided around by the cloak to emerge traveling in the same direction *as if* it had passed through the empty volume of space. An observer concludes that the secure volume is empty, but we are free to hide an object in the secure space. An alternative scheme has been recently investigated for the concealment of objects, (16) but relies on a specific knowledge of the shape and the material properties of the object being hidden. The electromagnetic cloak and the object concealed thus form a composite whose scattering properties can be reduced in the lowest order approximation: if the object changes the cloak must change too. In the scheme described here, an arbitrary object may be hidden because it remains untouched by external radiation. The method leads, in principle, to a perfect electromagnetic shield, excluding both propagating waves as well as near-fields from the concealed region.

For simplicity we choose the hidden object to be a sphere of radius $R_1$ and the cloaking region to be contained within the annulus $R_1 < r < R_2$. A simple transformation that achieves the desired result can be found by taking all fields in the region $r < R_2$ and compressing them into the region $R_1 < r < R_2$,

$$r' = R_1 + r(R_2 - R_1)/R_2,$$
$$\theta' = \theta, \quad (6)$$
$$\phi' = \phi$$

Applying the transformation rules (see the appendix), gives the following values:

for $r < R_1$: $\varepsilon', \mu'$ are free to take any value without restriction and do not contribute to electromagnetic scattering,

for $R_1 < r < R_2$:

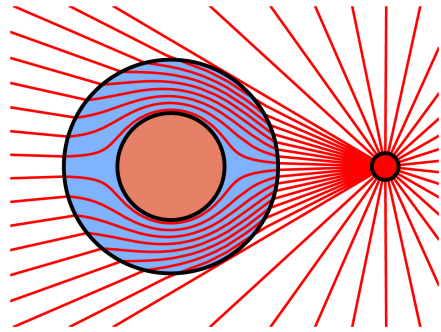

Fig. 3 A point charge located near the cloaked sphere. We assume that $R_2 \ll \lambda$, the near field limit, and plot the electric displacement field. The field is excluded from the cloaked region, but emerges from the cloaking sphere undisturbed. Note we plot field lines closer together near the sphere in order to emphasize the screening effect.

$$\varepsilon'_{r'} = \mu'_{r'} = \frac{R_2}{R_2 - R_1} \frac{(r' - R_1)^2}{r'^2},$$
$$\varepsilon'_{\theta'} = \mu'_{\theta'} = \frac{R_2}{R_2 - R_1}, \quad (7)$$
$$\varepsilon'_{\phi'} = \mu'_{\phi'} = \frac{R_2}{R_2 - R_1}$$

for $r > R_2$:
$$\varepsilon'_{r'} = \mu'_{r'} = \varepsilon'_{\theta'} = \mu'_{\theta'} = \varepsilon'_{\phi'} = \mu'_{\phi'} = 1 \quad (8)$$

We stress that this prescription will exclude *all* fields from the central region. Conversely no fields may escape from this region.

For purposes of illustration suppose that $R_2 \gg \lambda$ where $\lambda$ is the wavelength so that we can use the ray approximation to plot the Poynting vector. If our system is then exposed to a source of radiation at infinity we can perform the ray tracing exercise shown in Fig. 2. Rays in this figure result from numerical integration of a set of Hamilton's equations obtained by taking the geometric limit of Maxwell's equations with anisotropic, inhomogeneous media. This integration provides an independent confirmation that the configuration specified by (6) and (7) excludes rays from the interior region.

Alternatively if $R_2 \ll \lambda$ and we locate a point charge nearby, the electrostatic (or magnetostatic) approximation applies. A plot of the local electrostatic displacement field is shown in Fig. 3.

Next we discuss the characteristics of the cloaking material. There is an unavoidable singularity in the ray tracing, as can be seen by considering a ray headed directly towards the centre of the sphere (Fig. 2). This ray does not know whether to be deviated up or down, left or right. Neighboring rays are bent around in tighter and tighter arcs the closer to the critical ray they are. This in turn implies very rapid changes in $\varepsilon'$ and $\mu'$, as sensed by the ray. These rapid changes are due (in a self-consistent way) to the tight turn of the ray and the anisotropy of $\varepsilon'$ and $\mu'$. Anisotropy of the medium is necessary because we have compressed space anisotropically.

Although anisotropy and even continuous variation of the parameters is not a problem for metamaterials (18, 19, 20), achieving very large or very small values of $\varepsilon'$ and $\mu'$ can be. In practice, cloaking will be imperfect to the degree that we fail to satisfy equation (7). However, very considerable reductions in the cross section of the object can be achieved.

A further issue is whether the cloaking effect is broad band or specific to a single frequency. In the example we have given, the effect is only achieved at one frequency. This can easily be seen from the ray picture (Fig. 2). Each of the rays intersecting the large sphere is required to follow a curved and therefore longer trajectory than it would have done in free space, and yet we are requiring the ray to arrive on the far side of the sphere with the same phase. This implies a phase velocity greater that the velocity of light in vacuum which violates no physical law. However if we also require absence of dispersion, the group and phase velocities will be identical, and the group velocity can never exceed the velocity of light. Hence in this instance the cloaking parameters must disperse with frequency, and therefore can only be fully effective at a single frequency. We mention in passing that the group velocity may sometimes exceed the velocity of light (*21*) but only in the presence of strong dispersion. On the other hand if the system is embedded in a medium having a large refractive index, dispersion may in principle be avoided and the cloaking operate over a broad bandwidth.

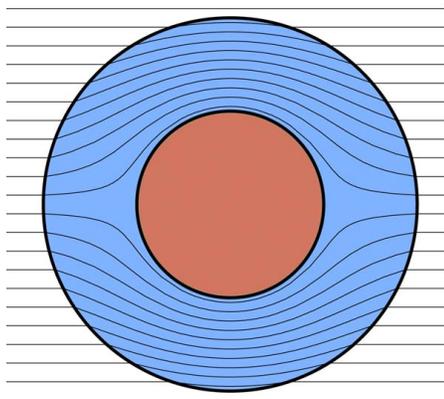
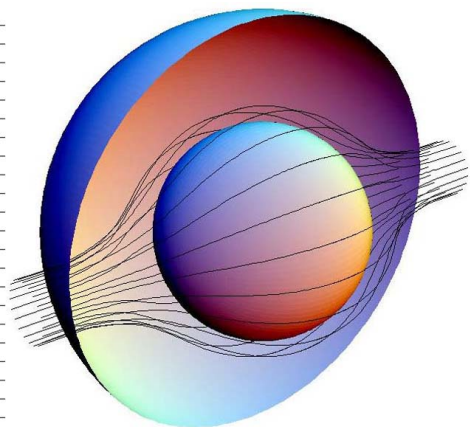

Fig. 2 A ray tracing program has been used to calculate ray trajectories in the cloak assuming that $R_2 \gg \lambda$. The rays essentially following the Poynting vector. **Left**: a 2D cross section of rays striking our system, diverted within the annulus of cloaking material contained within $R_1 < r < R_2$ to emerge on the far side undeviated from their original course. **Right**: a 3D view of the same process.

In conclusion, we have shown how electromagnetic fields can be dragged into almost any desired configuration. The distortion of the fields is represented as a coordinate transformation, which is then used to generate values of electrical permittivity and magnetic permeability ensuring that Maxwell's equations are still satisfied. The new concept of metamaterials is invoked making realization of these designs a practical possibility.

**References and Notes**
1. J. B. Pendry, A. J. Holden, W. J. Stewart, I. Youngs, *Phys. Rev. Lett.*, **76**, 4773 (1996).
2. J. B. Pendry, A. J. Holden, D. J. Robbins and W. J. Stewart, *IEEE Trans. Micr. Theory and Techniques*, **47**, 2075 (1999).
3. V. G. Veselago, *Soviet Physics USPEKI* **10**, 509 (1968).
4. D. R. Smith, W. J. Padilla, D. C. Vier, S. C. Nemat-Nasser, S. Schultz, *Phys. Rev. Lett.*, **84**, 4184 (2000).
5. R. A. Shelby, D. R. Smith, S. Schultz, *Science* **292**, 77 (2001).
6. A. A. Houck, J. B. Brock, I. L Chuang, *Phys. Rev. Lett.,* **90**, 137401 (2003).
7. A. Grbic, G. V. Eleftheriades, *Phys. Rev. Lett.*, **92**, 117403 (2004).
8. V. M. Shalaev, W. Cai, Uday K. Chettiar, H. K. Yuan, A. K. Sarychev, V. P. Drachev, and A. V. Kildishev, *Opt. Lett.*, **30**, 3356 (2005).
9. D.R. Smith, J.B. Pendry, M.C.K. Wiltshire, *Science,* **305,** 788 (2004)
10. E. Cubukcu, K. Aydin, E. Ozbay, S. Foteinopoulou, C. M. Soukoulis, *Nature*, **423**, 604 (2003).
11. E. Cubukcu, K. Aydin, E. Ozbay, S. Foteinopoulou, C. M. Soukoulis, *Phys. Rev. Lett.*, **91**, 207401 (2003).
12. T. J. Yen, W. J. Padilla, N. Fang, D. C. Vier, D. R. Smith, J. B. Pendry, D. N. Basov, and X. Zhang, *Science* **303**, 1494 (2004).
13. S. Linden, C. Enkrich, M. Wegener, J. F. Zhou, T. Koschny, and C. M. Soukoulis, *Science* **306**, 1351 (2004).
14. A. J. Ward and J. B. Pendry, *Journal of Modern Optics*, **43** 773 (1996).
15. U. Leonhardt, *IEEE Journal of Selected Topics in Quantum Electronics*, **9**, 102 (2003).
16. A. Alu, N. Engheta, *Phys. Rev.* **E95**, 016623 (2005).
17. J.-P. Berenger, *Journal of Computational. Physics*, 114, 185, (1994).
18. D. R. Smith, J. J. Mock, A. F. Starr, D. Schurig, *Phys. Rev. E* **71**, 036617 (2005).
19. T. Driscoll, D. N. Basov, A. F. Starr, P. M. Rye, S. Nemat-Nasser, D. Schurig, D. R. Smith, *Appl. Phys. Lett.* **88**, 081101 (2006).
20. R. B. Greegor, C. G. Parazzoli, J. A. Nielsen, M. A. Thompson, M. H. Tanielian, D. R. Smith, *Appl. Phys. Lett.* **87**, 091114 (2005).
21. R. Y. Chiao and P. W. Milonni, *Optics and Photonics News*, June (2002).
22. JBP thanks the EPSRC for a Senior Fellowship, the EC under project FP6-NMP4-CT-2003-505699, DoD/ONR MURI grant N00014-01-1-0803, DoD/ONR grant N00014-05-1-0861, and the EC Information Societies Technology (IST) programme Development and Analysis of Left-Handed Materials (DALHM), Project number: IST-2001-35511, for financial support. David Schurig would like to acknowledge support from the IC Postdoctoral Fellowship Program.